\documentclass[prd,twocolumn,
tightenlines,%
superscriptaddress,nofootinbib,notitlepage,preprintnumbers]{revtex4}
\usepackage{amssymb}
\usepackage{amsmath}
\usepackage{hyperref}
\usepackage{graphicx}

\begin{document}

\title{
Ghosts without runaway
}

\author{C\'edric~Deffayet}
\email{cedric.deffayet@iap.fr}
\affiliation{$\mathcal{G}\mathbb{R}\varepsilon{\mathbb{C}}\mathcal{O}$, Institut d'Astrophysique de Paris, UMR 7095, CNRS, Sorbonne Universit{\'e}, 98\textsuperscript{bis} boulevard Arago, 75014 Paris, France}
\affiliation{IHES, Le Bois-Marie, 35 route de Chartres, F-91440 Bures-sur-Yvette, France}

\author{Shinji Mukohyama} 
\email{shinji.mukohyama@yukawa.kyoto-u.ac.jp}
\affiliation{Center for Gravitational Physics, Yukawa Institute for Theoretical Physics,
Kyoto University, 606-8502, Kyoto, Japan}
\affiliation{Kavli Institute for the Physics and Mathematics of the Universe (WPI),
The University of Tokyo, Kashiwa, Chiba 277-8583, Japan}

\author{Alexander Vikman} 
\email{vikman@fzu.cz}
\affiliation{CEICO--Central European Institute for Cosmology and Fundamental Physics, FZU--Institute of Physics of the Czech Academy of Sciences, Na Slovance 1999/2, 18221 Prague 8, Czech Republic}

\preprint{YITP-21-82, IPMU21-0052}

\begin{abstract}
We present a simple class of mechanical models where a canonical degree of freedom interacts with another one with a negative kinetic term, i.e. with a ghost. We prove analytically that the classical motion of the system is completely stable for all initial conditions, notwithstanding that the conserved Hamiltonian is unbounded from below and above. This is fully supported by numerical computations. Systems with negative kinetic terms often appear in modern cosmology, quantum gravity and high energy physics and are usually deemed as unstable. Our result demonstrates that for mechanical systems this common lore can be too naive and that living with ghosts can be stable. 
\end{abstract}

\date{\today}

\maketitle

%{\it Introduction.} 
There are various reasons to be interested in field theories, or their simpler mechanical counterparts, where one degree of freedom has a ghostly nature, i.e. a negative kinetic term, while the other degrees of freedom have the usual positive-definite kinetic terms. As first shown by Ostrogradsky \cite{Ostrogradsky:1850fid} (for recent reviews see e.g. \cite{Woodard:2015zca,Bruneton:2007si}), such a situation appears in the Hamiltonian formulation of theories with higher derivative interactions, while such interactions have in turn interesting properties to regulate field theories in the UV \cite{Lee:1970iw,Stelle:1976gc,Grinstein:2007mp}.
A ghost copy of the standard model was also invoked in attempts to solve the cosmological constant problem \cite{Linde:1984ir,Linde:1988ws,Kaplan:2005rr}.
More recently, ghostly theories have also emerged in cosmology to describe nonstandard dynamics of the universe including bouncing cosmologies (see e.g. \cite{Brandenberger:2016vhg}) and dark energy with a phantom equation of state \cite{Caldwell:1999ew} which is still allowed by the latest cosmological data (e.g. \cite{Planck:2018vyg}). 
 Related proposals can also address the Hubble tension problem, see e.g. \cite{DiValentino:2021izs}. Sometimes such nonstandard dynamics can be obtained in ghost free theories (see e.g. \cite{Rubakov:2014jja}) where, however, the Hamiltonians are necessarily unbounded from below \cite{Sawicki:2012pz}. This implies a potential vulnerability to nonlinear instabilities.

The standard lore is of course that theories with ghosts and/or energies unbounded from below are unstable and as such problematic, even though various authors have advocated differently \cite{Lee:1970iw,Hawking:2001yt,Grinstein:2007mp,Garriga:2012pk,Salvio:2014soa,Smilga:2017arl,Anselmi:2018kgz,Donoghue:2021eto}. The instability inherent to ghostly models, usually dramatic at the quantum field theory level (see e.g \cite{Cline:2003gs}), is already seen at the classical level in the associated Hamiltonian motion. More specifically it can be linked to interactions between a ghostly degree of freedom and one of positive energy, as one isolated ghost would be stable. There again it is usually believed that {\it any} such interactions would generically lead to catastrophic trajectories with divergences or runaway instabilities already at the classical level.

However, there exist scarce studies indicating that this could be more subtle. Indeed, the well known KAM theorem (see e.g. \cite{1978mmcm.book.....A}) 
opens the way to stable motions in systems with a specific ghost-positive energy degree of freedom interaction and a restricted set of initial conditions, so called ``islands of stability'' \cite{Smilga:2004cy}. 
An analytic study of such a situation has been carried \cite{Pagani:1987ue}, showing that, for a specific model, there exist stable motions in the vicinity of one particular point in phase space. Some numerical works have reached similar conclusions in a restricted set of models \cite{Carroll:2003st, Smilga:2004cy, Pavsic:2016ykq, Smilga:2017arl}. On the other hand ref.~\cite{Pavsic:2013noa} found numerically, for one specific model, stable motions for all initial conditions.
 However, all these findings are either based on numerical integrations and, as such, are not fully conclusive  (as they cannot cover all the Hamiltonian trajectories) and/or 
only yields islands of stability, and not global stability, all this being true in addition at best for a very restricted set of interactions.

Here we propose a new look on these issues and present a large set of models, with a ghost in interaction with a positive energy degree of freedom, which have stable classical motions for all initial values. This stability is proven analytically. 

The model considered here is defined by a particular interaction potential $V_I\left(x,y\right)$ between a normal harmonic oscillator $x$ and a ghost oscillator $y$ of the same frequency. It has a Hamiltonian
\begin{equation} \label{Hamiltonian}
H=\frac{1}{2}\left(p_{x}^{2}+x^{2}\right)-\frac{1}{2}\left(p_{y}^{2}+y^{2}\right)+V_I\left(x,y\right)\,,
\end{equation}
with  $V_I\left(x,y\right)$ given by 
\begin{equation} \label{Potential}
V_{I}\left(x,y\right)=\lambda \left(\left(x^{2}-y^{2}-1\right)^{2}+4x^{2}\right)^{-1/2}\,,
\end{equation}
where $\lambda$ is the coupling constant. 
%%%%%%%%Fig_1%%%%%%%%
\begin{figure}[tb!]
  \begin{center}
   \includegraphics[width=\columnwidth,angle=0]{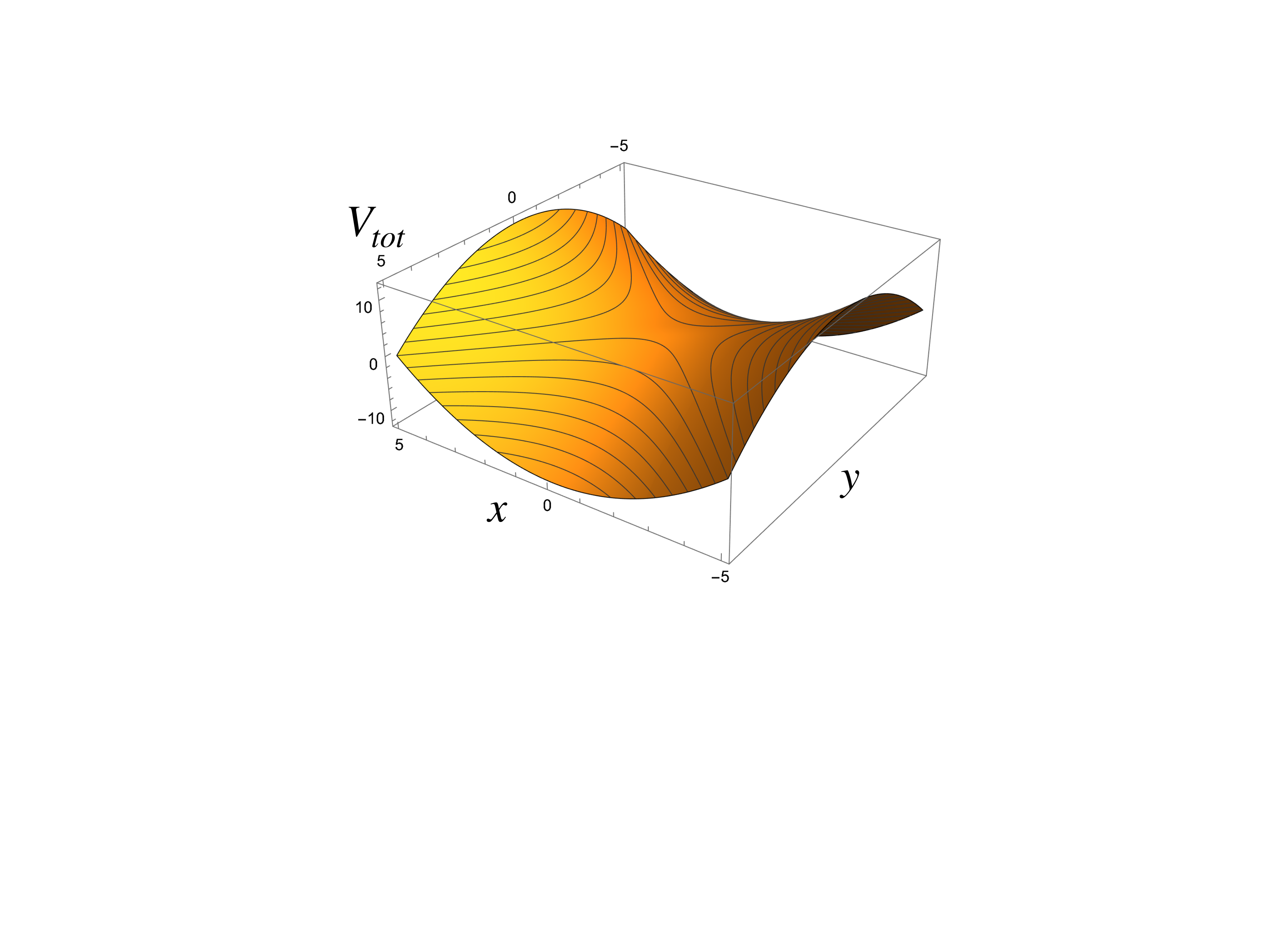}
  \caption{Total potential energy is plotted for coupling constant $\lambda=1/3$ in the interaction potential~(\ref{Potential}).}
  \label{Potential_Fig}
  \end{center}
\end{figure}
%%%%%%%%Fig_1%%%%%%%%
The model is well defined for all values of $x$ and $y$. Indeed, the expression under the square root in \eqref{Potential} is positive definite so that the interaction potential $V_I$ is always smooth and finite. The total potential energy $V_{tot}$, see Fig.~\ref{Potential_Fig}, and the total Hamiltonian are unbounded from above and from below. However, the interaction potential $V_I$ is bounded as 
\begin{equation} \label{boundV}
0 < V_I\left(x,y\right) \lambda^{-1} \leq 1\,.
\end{equation} 
Expanding the total potential energy in $x$ and $y$ around the origin yields: 
\begin{equation} \label{expandV}
V_{tot}=\frac{\omega_{x}^{2}}{2}x^{2}-\frac{\omega_{y}^{2}}{2}y^{2}+\lambda \left(x^{4}+4y^{2}x^{2}+y^{4}\right)+... \,,
\end{equation}
where the frequencies are corrected by the interaction as 
\begin{equation} \label{omega}
\omega_{x}^{2}=1-2\lambda\,,\quad\text{and}\quad\omega_{y}^{2}=1+2\lambda\,,
\end{equation}
and where we neglected an irrelevant constant and terms of total power higher than $4$. 
Thus, for $|\lambda|<1/2$, both oscillators are linearly stable around the origin~\footnote{For $\lambda>1/2$, the system exhibits tachyonic instability around the origin, but is locally stable around the saddles of the total potential at $(x, y)=(\pm(\sqrt{2\lambda}-1)^{1/2}, 0)$. For $\lambda<-1/2$, on the the hand, the system is locally stable around saddles at $(x, y)=(0, \pm(\sqrt{-2\lambda}-1)^{1/2})$.}. However, we will not here restrict ourselves to motions staying in the vicinity of some point in phase space, and which could possibly be described by perturbation theory. Rather we will prove a stability theorem valid for all initial conditions and Hamiltonian motions. To that hand it is crucial to note that the model defined by \eqref{Hamiltonian}-\eqref{Potential} is integrable: in addition to the conserved Hamiltonian $H$, it has a first integral 
\begin{equation}
  C =K^{2}+\left(p_{x}^{2}+x^{2}\right)
   -\left(x^{2}-y^{2}-1\right)V_{I}\left(x,y\right)\,,\label{FirstIntegral}
\end{equation}
where $K=p_{y}x+p_{x}y$ is the momentum of hyperbolic rotations (boosts) in $\left(x,y\right)$ plane. 
One can explicitly check that the above quantity $C$ is conserved by the Hamiltonian motion. However, we also note that the above model can be obtained from a class of integrable models obtained by Darboux in 1901 \cite{Darboux} with two positive energy degrees of freedom $x$ and $\tilde y$, using the complex canonical transformations 
\begin{equation}
\label{complex_trans}
y=i \tilde{y}\,,\quad\text{and}\quad p_y=-i \tilde{p}_y\,,
\end{equation}
which in our case not only preserve the Hamiltonian motion but also keep both $H$ and $C$ real. 
It is useful to introduce the sum of absolute values of the energies of both oscillators and the square of $K$
\begin{equation}\label{EnergySum}
\Sigma=\left(p_{y}x+p_{x}y\right){}^{2}+\frac{1}{2}\left(p_{x}^{2}+x^{2}\right)+\frac{1}{2}\left(p_{y}^{2}+y^{2}\right)\,.
\end{equation}
Each term in $\Sigma$ is manifestly non-negative and corresponds to a first integral of the system \eqref{Hamiltonian} without interaction, i.e. for $\lambda=0$. 
The distance from a point in the phase space $\xi=\left(x,y,p_{x},p_{y}\right)$ to the origin (the euclidean norm of the state) is always bounded:
$\left|\xi\right|^{2}\leq2\Sigma$. The following difference $\mathcal{E}$ between two integrals of motion is useful:
\begin{equation}
\label{Difference}
\mathcal{E}=C-H=\Sigma+\left(y^{2}-x^{2}\right)V_{I}\left(x,y\right)\,.
\end{equation}
The second term in the last equality is bounded in the stripe
\begin{equation}
\label{Bound}
-|\lambda| \leq \left(y^{2}-x^{2}\right)V_{I}\left(x,y\right) \leq |\lambda|\,.
\end{equation}
As a consequence, one gets that at all times 
\begin{equation}
\label{Bound_on_E}
\Sigma-|\lambda| \leq \mathcal{E} \leq \Sigma+|\lambda|\,.  
\end{equation}
Applying this inequality at two different times $t_a$ and $t_b$, and using the conservation of $\mathcal{E}$ we get that (the index $a,b$ refering to the corresponding time)
\begin{equation}
\label{Onion_Belt}
\Sigma_{a}-2|\lambda| \leq \Sigma_{b} \leq \Sigma_{a}+2|\lambda|\,,
\end{equation}
In particular, the last inequality implies that 
\begin{equation}
\label{Bound_Radius}
\left|\xi_{b}\right|^{2} \leq  \left|\xi_{a}\right|^{2}+2K_{a}^{2}+4|\lambda|\,,
\end{equation}
using $2\Sigma = 2K^2+|\xi|^2 \geq |\xi|^2$. 
Hence, the motion is confined inside of a sphere whose radius is fixed by the (initial) data at $t_a$. 
This completes the proof that the motion of our system is always bounded and shows no runaway for all values of the initial data and coupling constant $\lambda$. We stress that this 
also holds for parameters yielding ``tachyonic'' negative $\omega_{x}^{2}$ or/and $\omega_{y}^{2}$ corresponding to an unstable origin.

\begin{figure}[tb!]
  \begin{center}
   \includegraphics[width=\columnwidth,angle=0]{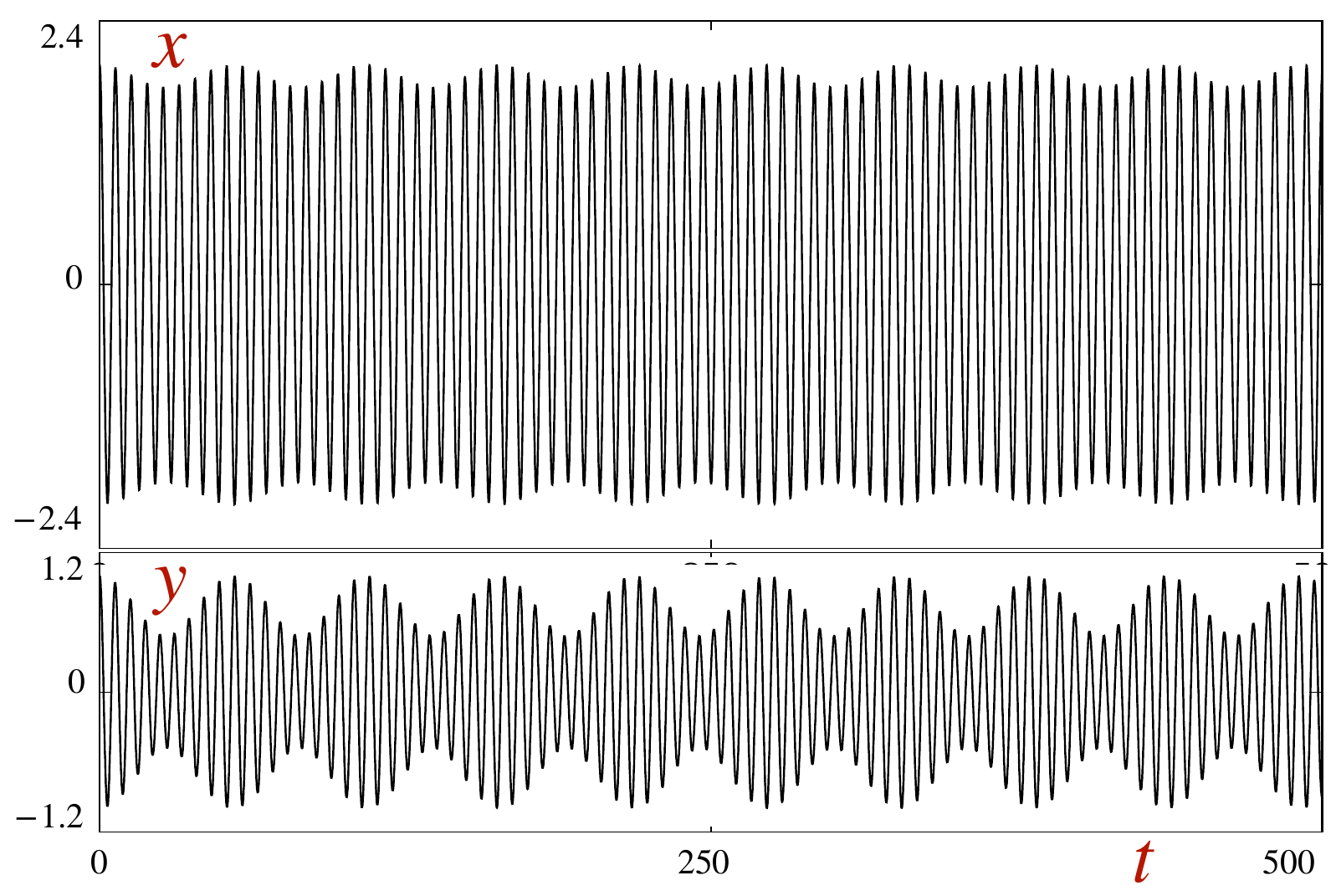}
  \caption{The plot of $x(t)$ and $y(t)$.}
  \label{x+y_Fig}
  \end{center}
\end{figure}

\begin{figure}[tb!]
  \begin{center}
   \includegraphics[width=\columnwidth,angle=0]{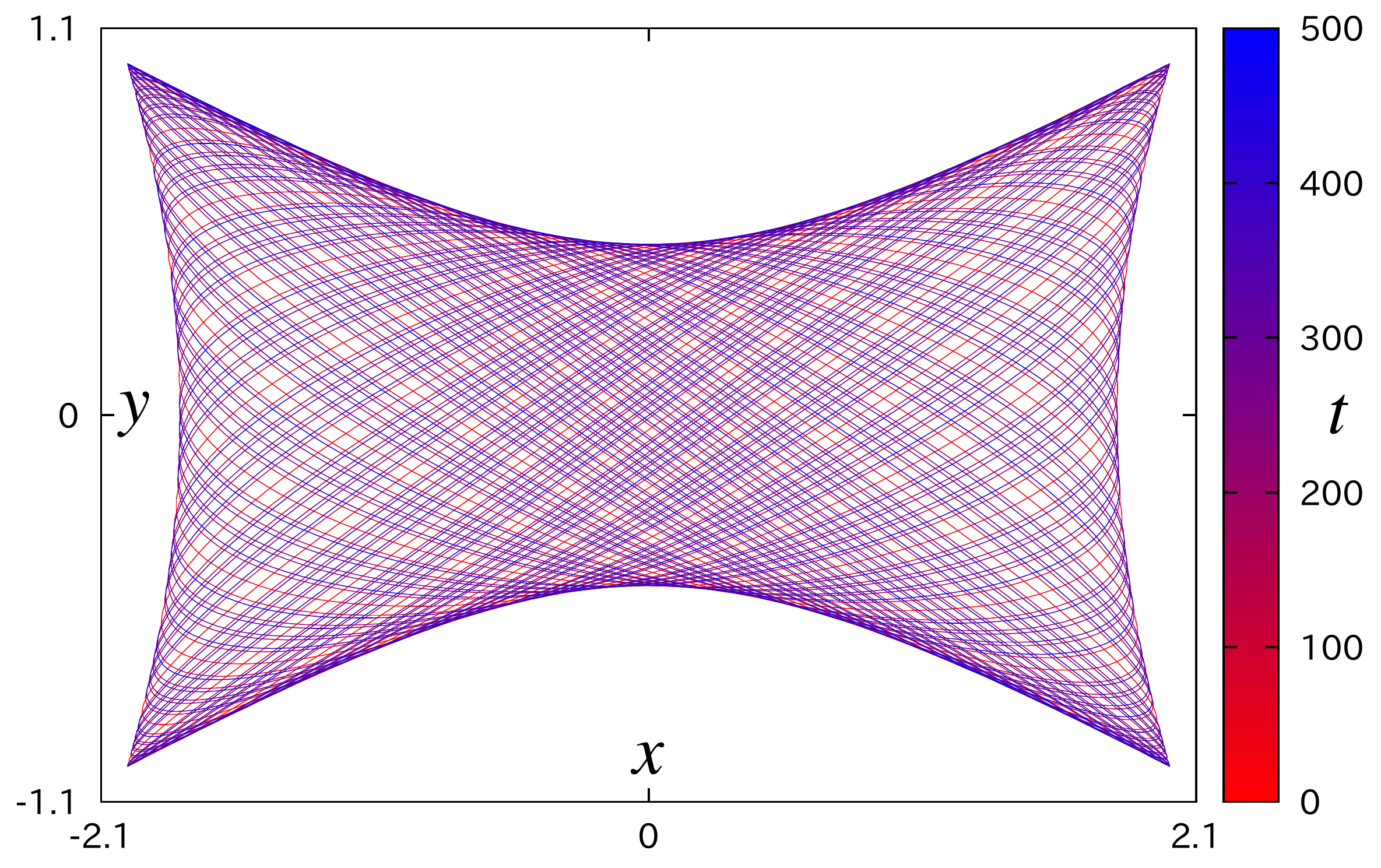}
  \caption{The projection of the trajectory onto the $xy$ plane. The color represents the value of $t$.}
  \label{xy_Fig}
  \end{center}
\end{figure}

\begin{figure}[tb!]
  \begin{center}
   \includegraphics[width=\columnwidth,angle=0]{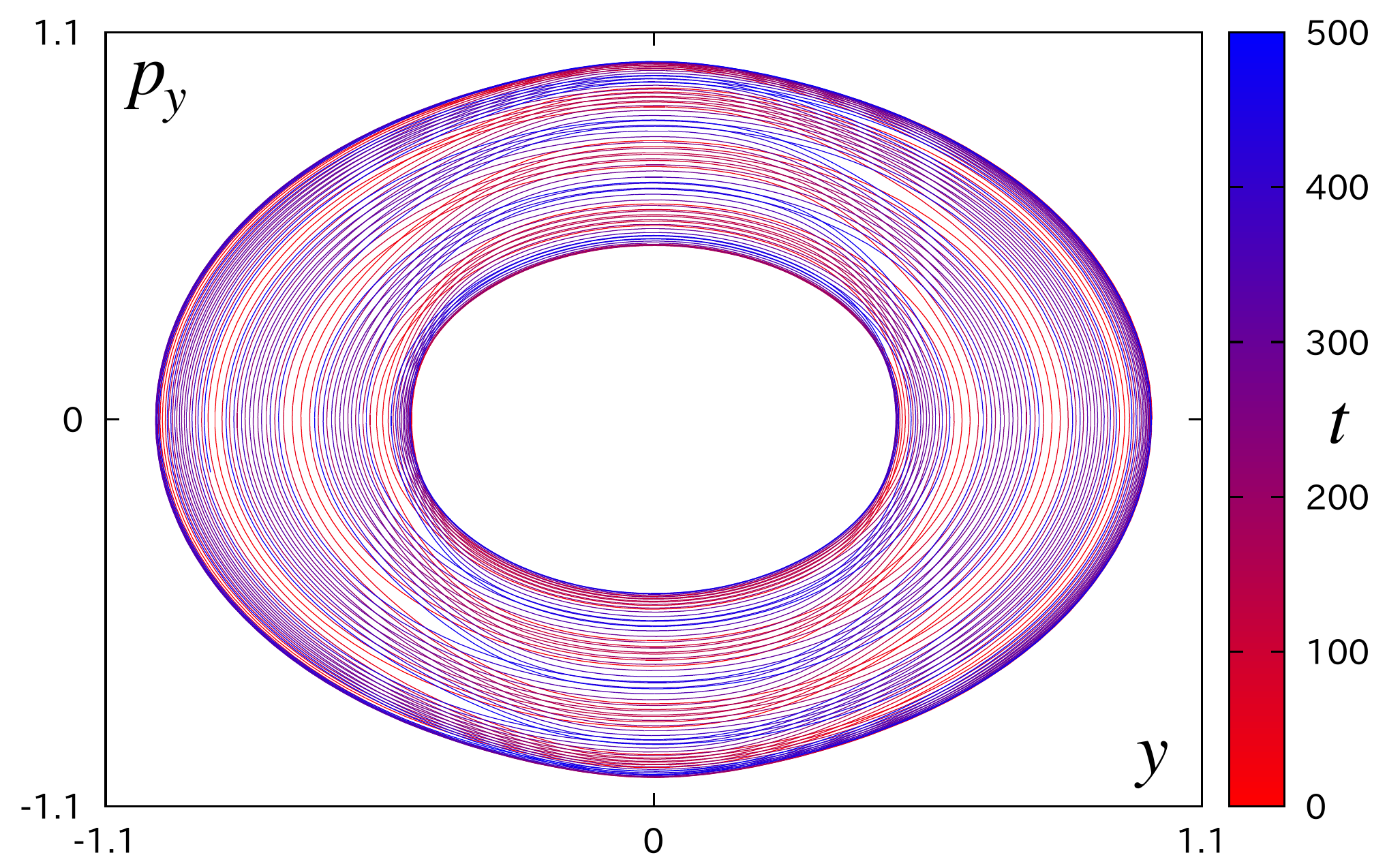}
  \caption{The projection of the trajectory onto the $y p_y$ plane. The color represents the value of $t$.}
  \label{Ellipse}
  \end{center}
\end{figure}

We now show the result of numerical integration of the system (\ref{Hamiltonian})-(\ref{Potential}) with $\lambda=1/3$. The Hamiltonian equation of motion is directly solved numerically from the initial time $t=0$ to the final time $t=500$ with the initial condition $(x(0), y(0), p_{x}(0), p_{y}(0))=(2, 1, 0, 0)$. Numerical errors in $H$ and $C$ remain of order $10^{-13}$. Fig.~\ref{x+y_Fig}  shows the behaviors of $x(t)$ and $y(t)$. Each of them stably oscillates with some modulation induced by the interaction between them. Figs.~\ref{xy_Fig} and~\ref{Ellipse} show the projection of the trajectory onto the $xy$ and $yp_y$  planes respectively, the color represents the value of $t$. 
%%%%%%%%Fig_2%%%%%%%%
%%%%%%%%Fig_2%%%%%%%%

%%%%%%%%Fig_3%%%%%%%%
%%%%%%%%Fig_3%%%%%%%%

%%%%%%%%Fig_4%%%%%%%%
%%%%%%%%Fig_4%%%%%%%%
Note that the trajectories close to the origin can also be analyzed more precisely. First of all, from the first inequality in (\ref{Onion_Belt}), one obtains
\begin{equation}\label{Bound_upper}
\left|\xi_{a}\right|^{2}-2K_{b}^{2}-4|\lambda| \leq \left|\xi_{b}\right|^{2}\,.
\end{equation}
Close to the origin of the phase space, $K^2$ is higher order than $\left|\xi\right|^2$ and can be neglected. Thus, in this case, we conclude that, for $4|\lambda|<|\xi|^2\ll 1$, the trajectories are located in the spherical shell which is $4|\lambda|/|\xi_{a}|$ thin:
\begin{equation}\label{Bound_lower}
 \left|\xi_{a}\right|^{2}-4|\lambda|
  \lesssim \left|\xi_{b}\right|^{2}
 \lesssim \left|\xi_{a}\right|^{2}+4|\lambda|\,, \nonumber
\end{equation}
where we omitted $\mathcal{O}(|\xi|^4)$ terms. 
Close to the origin one can use \eqref{boundV} to refine \eqref{Bound_on_E}, so that for $\lambda\left(y^{2}-x^{2}\right)<0$
\begin{equation}\label{Refine_Neg} 
\Sigma+\lambda\left(y^{2}-x^{2}\right) \leq \mathcal{E} \leq \Sigma\,,
\end{equation}
while for $\lambda\left(y^{2}-x^{2}\right)\geq0$ the limits flip and yield $\mathcal{E}>0$ (except at the origin). 
However, 
\begin{equation}\label{lower_bound} 
\Sigma+\lambda\left(y^{2}-x^{2}\right)=K{}^{2}+\frac{1}{2}\left(p_{x}^{2}+p_{y}^{2}+\omega_{x}^{2}x^{2}+\omega_{y}^{2}y^{2}\right)\,,
\end{equation}
where $\omega_{x}$ and $\omega_{y}$ are given by \eqref{omega}.  For $|\lambda| < 1/2$ both  $\omega_{x}^2$ and $\omega_{y}^2$ are positive, which results in $\mathcal{E} > 0$, except at the origin where $\mathcal{E}=0$. Thus, $\mathcal{E}$ being in addition conserved, satisfies all requirements of a Lyapunov function and guaranties the stability of the origin for $|\lambda| < 1/2$.

A stable (and integrable) motion such as the one discovered here can of course also be observed for a system of a ghost non interacting with a positive energy degree of freedom. Hence a legitimate question is whether one could transform the model considered here into such a system, i.e. find a canonical transformation which kills all the interactions. One can show, at least order by order, that this is not possible. To that end one can use a theorem given in \cite{1978mmcm.book.....A}, which in the present context amounts to state that, using the new variables 
defined by $z_x = p_x + i x$ and $\bar{z}_x = p_x - i x$, any interaction of the form 
\begin{equation} \label{monomial}
z_x^{{\alpha}_x} \bar{z}_x^{{\beta}_x} z_y^{{\alpha}_y} \bar{z}_y^{{\beta}_y}
\end{equation}
can be removed by a suitable canonical transformation except in the cases where $\alpha_x = \beta_x$ and simultaneously $\alpha_y = \beta_y$. This holds true in the so-called non resonant case which includes the case considered here where the ratio $\omega_{x}/\omega_{y} = \sqrt{(1-2\lambda)/(1+2\lambda)}$ is generically irrational. In our case, it is easy to see that each term $x^4$, $x^2 y^2$ and $y^4$ appearing at order $4$ in the expansion 
of the potential (\ref{expandV}) contains one and only one monomial which cannot be removed, respectively given by the distinct monomials 
$z_x^2 \bar{z}_x^2$, $z_x \bar{z}_x z_y \bar{z}_y$ and $z_y^2 \bar{z}_y^2$. Hence, we conclude that it is not possible to fully remove the quartic interaction of our model via a canonical transformation that keeps the quadratic part of the Hamiltonian.

We note further that the model considered here and defined by the Hamiltonian (\ref{Hamiltonian}) can easily, at least locally, be rewritten as a higher derivative theory for a single degree of freedom $q$. To that hand, one inverts Ostrogradsky procedure and ends up with an equivalent Lagrangian $L$ given by (where a dot means a time derivative)
\begin{eqnarray}
L(q,\ddot{q}) = \left(\ddot{q}+q\right) \left(2 p_2 + (2 p_2)^{-1}\right)\,,
\end{eqnarray} 
where $p_2\equiv p_2\left(q,\ddot{q}\right)$ is solution of the equation 
\begin{equation}
(\ddot{q}+q)\sqrt{2 q^2+1} = - 2 \lambda p_2 (2 p_2^2+1)^{-3/2}
\end{equation} 
and is the Legendre conjugate variable to $\ddot{q}$ in the Lagrangian $L(q,\ddot{q})$ (i.e. one has $p_2 = \partial L/\partial \ddot{q}$).

Last, we underline that the above model (\ref{Hamiltonian}) is not unique. It is part of a larger family of models with the Hamiltonian $H = p_{x}^{2}/2 - p_{y}^{2}/2 + V$, where 
\begin{equation}
 V =  a U + b W+ c U W\,,
\end{equation}
where 
\begin{equation}
U = d -x^2 + y^2\,, \quad 
W = \left(U^2 + 4 d x^2\right)^{-1/2}\,,
\end{equation}
and $a,b,c,d$ are arbitrary constants satisfying $a <0$, $c \leq 0$ and $d >0$. These models are all integrable, with a motion whose stable nature can be proven analytically along the line above, with a Hamiltonian unbounded below and above and a ghost coupling to a positive energy degree of freedom \cite{inprep}.

%{\it Conclusions.} 
We have presented an example of classical models where a subsystem with positive energy unbounded from above interacts with another subsystem with negative energy unbounded from below. Yet the dynamics is such that the negative energy is locked and cannot be exploited to further increase the positive energy of the other subsystem. Hence, there are no runaway solutions in the whole phase space.  Moreover, we have shown the Lyapunov stability of the origin in the model (\ref{Hamiltonian}) (for $|\lambda|<1/2$), while our numerical investigations indicate that such a stability exists more widely on phase space \cite{inprep}. Note also that as the system is integrable, the KAM theorem should allow existence of ``islands of stability'' for a large class of non integrable interactions around the considered models. We stress however that the integrability, which plays an important role in our proof, does not {\it per se} guarantee the absence of runaway solutions \cite{inprep}. It would be very interesting to understand quantum mechanical description of such systems and their generalisation to continuum number of degrees of freedom.  Incidentally, the work reported here also points out the existence of a large set of integrable models where ghosts interact with a positive energy degree of freedom. These ghostly models can be obtained via known integrable models with two positive energy degrees of freedom and a complex canonical transformation of the form (\ref{complex_trans}) \cite{inprep}. To our knowledge, the only previously discussed example of such an ``integrable ghost'' (with a total of two degrees of freedom) is a very specific model given in \cite{Robert:2006nj} obtained from a supersymmetric field theory (see also \cite{Smilga:2020elp}).

It seems, that invoking interaction with ghosts may be rather innocent, at least in some cases. Thus, in these cases, it is not stability which precludes the existence of ghosts. One then expects to be able to find such stable interacting ghosts in some natural systems, in a wider context than the one mentioned in the introduction.

{\it Acknowledgments.} 
C.~D. thanks A.~Smilga and J.~F\'ejoz for discussions.
S.~M.'s work was supported in part by Japan Society for the Promotion of Science Grants-in-Aid for Scientific Research No. 17H02890, No. 17H06359, and by World Premier International Research Center Initiative, MEXT, Japan. A.~V. is supported in part by the European Regional Development Fund (ESIF/ERDF) and the Czech Ministry of Education, Youth and Sports (M\v SMT) through the Project CoGraDS- CZ.02.1.01/0.0/0.0/15 003/0000437. S.~M. and A.~V. collaboration is also supported by the Bilateral Czech-Japanese Mobility Plus Project JSPS-21-12.


\begin{thebibliography}{99}

%\cite{Ostrogradsky:1850fid}
\bibitem{Ostrogradsky:1850fid}
M.~Ostrogradsky,
%``M\'emoires sur les \'equations diff\'erentielles, relatives au probl\`eme des isop\'erim\`etres,''
Mem. Acad. St. Petersbourg \textbf{6} (1850) no.4, 385-517

%\cite{Woodard:2015zca}
\bibitem{Woodard:2015zca}
R.~P.~Woodard,
``Ostrogradsky's theorem on Hamiltonian instability,''
Scholarpedia \textbf{10} (2015) no.8, 32243
%doi:10.4249/scholarpedia.32243
%[arXiv:1506.02210 [hep-th]].

%\bibitem{Fecko}
%M.~Fecko, Ostrogradsky theorem (1850), Student Colloquium and School on Mathematical Physics, Star\'a Lesn\'a, Slovakia, 2015. 
%http://davinci.fmph.uniba.sk/~fecko1/referaty/stara\_lesna\_2015.pdf

%\cite{Bruneton:2007si}
\bibitem{Bruneton:2007si}
J.~P.~Bruneton and G.~Esposito-Farese,
%``Field-theoretical formulations of MOND-like gravity,''
Phys. Rev. D \textbf{76} (2007), 124012
[erratum: Phys. Rev. D \textbf{76} (2007), 129902]
%doi:10.1103/PhysRevD.76.129902
%[arXiv:0705.4043 [gr-qc]].

%\cite{Lee:1970iw}
\bibitem{Lee:1970iw}
T.~D.~Lee and G.~C.~Wick,
%``Finite Theory of Quantum Electrodynamics,''
Phys. Rev. D \textbf{2} (1970), 1033-1048
%doi:10.1103/PhysRevD.2.1033

%\bibitem{Woodard:2006nt}
%R.~P.~Woodard,
%``Avoiding dark energy with 1/r modifications of gravity,''
%Lect. Notes Phys. \textbf{720} (2007), 403-433
%doi:10.1007/978-3-540-71013-4\_14
%[arXiv:astro-ph/0601672 [astro-ph]].

%\cite{Stelle:1976gc}
\bibitem{Stelle:1976gc}
K.~S.~Stelle,
%``Renormalization of Higher Derivative Quantum Gravity,''
Phys. Rev. D \textbf{16} (1977), 953-969
%doi:10.1103/PhysRevD.16.953

%\cite{Grinstein:2007mp}
\bibitem{Grinstein:2007mp}
B.~Grinstein, D.~O'Connell and M.~B.~Wise,
%``The Lee-Wick standard model,''
Phys. Rev. D \textbf{77} (2008), 025012
%doi:10.1103/PhysRevD.77.025012
%[arXiv:0704.1845 [hep-ph]].

\bibitem{Linde:1984ir}
A.~D.~Linde,
%``The Inflationary Universe,''
Rept. Prog. Phys. \textbf{47} (1984), 925-986
%doi:10.1088/0034-4885/47/8/002

\bibitem{Linde:1988ws}
A.~D.~Linde,
%``The Universe Multiplication and the Cosmological Constant Problem,''
Phys. Lett. B \textbf{200} (1988), 272-274
%doi:10.1016/0370-2693(88)90770-8

\bibitem{Kaplan:2005rr}
D.~E.~Kaplan and R.~Sundrum,
%``A Symmetry for the cosmological constant,''
JHEP \textbf{07} (2006), 042
%doi:10.1088/1126-6708/2006/07/042
%[arXiv:hep-th/0505265 [hep-th]].

\bibitem{Brandenberger:2016vhg}
R.~Brandenberger and P.~Peter,
%``Bouncing Cosmologies: Progress and Problems,''
Found. Phys. \textbf{47} (2017) no.6, 797-850
%doi:10.1007/s10701-016-0057-0
%[arXiv:1603.05834 [hep-th]].

\bibitem{Caldwell:1999ew}
R.~R.~Caldwell,
%``A Phantom menace?,''
Phys. Lett. B \textbf{545} (2002), 23-29
%doi:10.1016/S0370-2693(02)02589-3
%[arXiv:astro-ph/9908168 [astro-ph]].

\bibitem{Planck:2018vyg}
N.~Aghanim \textit{et al.} [Planck],
%``Planck 2018 results. VI. Cosmological parameters,''
Astron. Astrophys. \textbf{641} (2020), A6
%doi:10.1051/0004-6361/201833910
%[arXiv:1807.06209 [astro-ph.CO]].

\bibitem{DiValentino:2021izs}
E.~Di Valentino\textit{et al.},
%``In the realm of the Hubble tension\textemdash{}a review of solutions,''
Class. Quant. Grav. \textbf{38} (2021) no.15, 153001
%doi:10.1088/1361-6382/ac086d
%[arXiv:2103.01183 [astro-ph.CO]].

\bibitem{Rubakov:2014jja}
V.~A.~Rubakov,
%``The Null Energy Condition and its violation,''
Phys. Usp. \textbf{57} (2014), 128-142
%doi:10.3367/UFNe.0184.201402b.0137
%[arXiv:1401.4024 [hep-th]].

\bibitem{Sawicki:2012pz}
I.~Sawicki and A.~Vikman,
%``Hidden Negative Energies in Strongly Accelerated Universes,''
Phys. Rev. D \textbf{87} (2013) no.6, 067301
%doi:10.1103/PhysRevD.87.067301
%[arXiv:1209.2961 [astro-ph.CO]].

%\cite{Hawking:2001yt}
\bibitem{Hawking:2001yt}
S.~W.~Hawking and T.~Hertog,
%``Living with ghosts,''
Phys. Rev. D \textbf{65} (2002), 103515
%doi:10.1103/PhysRevD.65.103515
%[arXiv:hep-th/0107088 [hep-th]].

\bibitem{Garriga:2012pk}
J.~Garriga and A.~Vilenkin,
%``Living with ghosts in Lorentz invariant theories,''
JCAP \textbf{01} (2013), 036
%doi:10.1088/1475-7516/2013/01/036
%[arXiv:1202.1239 [hep-th]].

%\cite{Salvio:2014soa}
\bibitem{Salvio:2014soa}
A.~Salvio and A.~Strumia,
%``Agravity,''
JHEP \textbf{06} (2014), 080
%doi:10.1007/JHEP06(2014)080
%[arXiv:1403.4226 [hep-ph]].

%\cite{Smilga:2017arl}
\bibitem{Smilga:2017arl}
A.~Smilga,
%``Classical and quantum dynamics of higher-derivative systems,''
Int. J. Mod. Phys. A \textbf{32} (2017) no.33, 1730025
%doi:10.1142/S0217751X17300253
%[arXiv:1710.11538 [hep-th]].

%\cite{Anselmi:2018kgz}
\bibitem{Anselmi:2018kgz}
D.~Anselmi,
%``Fakeons And Lee-Wick Models,''
JHEP \textbf{02} (2018), 141
%doi:10.1007/JHEP02(2018)141
%[arXiv:1801.00915 [hep-th]].

%\cite{Donoghue:2021eto}
\bibitem{Donoghue:2021eto}
J.~F.~Donoghue and G.~Menezes,
%``The Ostrogradsky instability can be overcome by quantum physics,''
[arXiv:2105.00898 [hep-th]].

%\cite{Cline:2003gs}
\bibitem{Cline:2003gs}
J.~M.~Cline, S.~Jeon and G.~D.~Moore,
%``The Phantom menaced: Constraints on low-energy effective ghosts,''
Phys. Rev. D \textbf{70} (2004), 043543
%doi:10.1103/PhysRevD.70.043543
%[arXiv:hep-ph/0311312 [hep-ph]].

\bibitem{1978mmcm.book.....A} Arnold, V.~I.\ 1978, Mathematical Methods of Classical Mechanics, Graduate texts in mathematics, New York: Springer, 1978.

%\cite{Smilga:2004cy}
\bibitem{Smilga:2004cy}
A.~V.~Smilga,
%``Benign versus malicious ghosts in higher-derivative theories,''
Nucl. Phys. B \textbf{706} (2005), 598-614

\bibitem{Pagani:1987ue}
E.~Pagani, G.~Tecchiolli and S.~Zerbini,
%``On the Problem of Stability for Higher Order Derivatives: Lagrangian Systems,''
Lett. Math. Phys. \textbf{14} (1987), 311
%doi:10.1007/BF00402140

%\cite{Carroll:2003st}
\bibitem{Carroll:2003st}
S.~M.~Carroll, M.~Hoffman and M.~Trodden,
%``Can the dark energy equation-of-state parameter $w$ be less than $−1$?,''
Phys. Rev. D \textbf{68} (2003), 023509
%doi:10.1103/PhysRevD.68.023509
%[arXiv:astro-ph/0301273 [astro-ph]].

%\cite{Pavsic:2016ykq}
\bibitem{Pavsic:2016ykq}
M.~Pav\v{s}i\v{c},
%``Pais\textendash{}Uhlenbeck oscillator and negative energies,''
Int. J. Geom. Meth. Mod. Phys. \textbf{13} (2016) no.09, 1630015
%doi:10.1142/S0219887816300154
%[arXiv:1607.06589 [gr-qc]]. 

%\cite{Pavsic:2013noa}
\bibitem{Pavsic:2013noa}
M.~Pav\v{s}i\v{c},
%``Stable Self-Interacting Pais-Uhlenbeck Oscillator,''
Mod. Phys. Lett. A \textbf{28} (2013), 1350165
%doi:10.1142/S0217732313501654
%[arXiv:1302.5257 [gr-qc]].

\bibitem{Darboux}
G.~Darboux, 
%% Sur un probl\`eme de m\'ecanique,
Archives N\'eerlandaises (ii) 6 (1901) 371.

%\cite{Hietarinta:1986tw}
%\bibitem{Hietarinta:1986tw}
%J.~Hietarinta,
%``Direct Methods for the Search of the Second Invariant,''
%Phys. Rept. \textbf{147} (1987), 87
%doi:10.1016/0370-1573(87)90089-5

\bibitem{inprep}
C.~Deffayet, S.~Mukohyama, A.~Vikman, in preparation.

%\cite{Robert:2006nj}
\bibitem{Robert:2006nj}
D.~Robert and A.~V.~Smilga,
%``Supersymmetry vs ghosts,''
J. Math. Phys. \textbf{49} (2008), 042104

%\cite{Smilga:2020elp}
\bibitem{Smilga:2020elp}
A.~V.~Smilga,
%``On exactly solvable ghost-ridden systems,''
Phys. Lett. A \textbf{389} (2021), 127104
%doi:10.1016/j.physleta.2020.127104
%[arXiv:2008.12966 [hep-th]].

\end{thebibliography}
\end{document}